\documentclass[prb,aps,showpacs,twocolumn,floatfix]{revtex4-1}
\usepackage{amsmath,amsfonts,amssymb,graphicx,bm}
\usepackage[english]{babel}
\newcommand{\bS}{{\bm S}}
\begin{document}
\title{Comment on "Quantum phase transition in the four-spin exchange
       antiferromagnet"}
\author{L. Isaev$^1$}
\author{G. Ortiz$^1$}
\author{J. Dukelsky$^2$}
\affiliation{$^1$Department of Physics, Indiana University, Bloomington IN
                 47405, USA \\
             $^2$Instituto de Estructura de la Materia - CSIC, Serrano 123,
                 28006 Madrid, Spain}
\begin{abstract}
 In a recent paper [Phys. Rev. {\bf B80}, 174403 (2009)] Kotov {\it et al.}
 studied  the paramagnetic-to-antiferromagnetic transition in the $J$-$Q$
 model. Their findings were claimed to be in ``fairly good agreement'' with
 previous quantum Monte-Carlo (QMC) results. In this Comment we show that the
 above claim is misleading and in reality their phase transition point 
 is not only far from the corresponding QMC value but also lies in a region of
 parameter space not yet explored in the literature. We also show that their
 reference dimer state is unstable against formation of a plaquette condensate,
 which could in part explain the large fluctuations they found. 
\end{abstract}
\maketitle

The $J$-$Q$ model, proposed by Sandvik \cite{Sandvik_2007}, is defined by the
Hamiltonian:
\begin{align}
 H&=J\sum_{\langle ij\rangle}\bS_i\bS_j-Q\sum_{\langle ijkl\rangle}\biggl(
 \bS_i\bS_j-\frac{1}{4}\biggr)\biggl(\bS_k\bS_l-\frac{1}{4}\biggr)
 \label{jq_hamiltonian} \\
 =&-\frac{NQ}{8}+\biggl(J+\frac{Q}{2}\biggr)\sum_{\langle ij\rangle}\bS_i
 \bS_j-Q\sum_{\langle ijkl\rangle}(\bS_i\bS_j)(\bS_k\bS_l), \nonumber
\end{align}
where $i$, $j$ \ldots denote sites in a 2D square lattice, $N$ is the number of
spins, and $\bS_i$ are spin-$1/2$ operators. The four-spin interaction $Q$ and
next-nearest neighbor exchange $J$ were originally assumed to be positive
\cite{Sandvik_2007}. For reasons that will become clear later, we will also
consider the sector with $J<0$ (but keeping $Q>0$). The model of Eq.
\eqref{jq_hamiltonian} with $J>0$ exhibits a quantum phase transition (QPT)
between the antiferromagnetic (AF) and paramagnetic singlet phases. However,
the location and nature of this transition as well as the structure of the
singlet phase are still debated. In particular, in Ref.
\onlinecite{Sandvik_2007} it was concluded, using QMC simulations, that the QPT
is $2^{\rm nd}$ order, occurs at $Q_c/J\approx25$ and is consistent with the
deconfined quantum criticality scenario \cite{Sachdev_2004}. In a later work
\cite{Isaev_2009} by the present authors, using a hierarchical mean-field (HMF)
approach, the location of this QPT was found at $Q_c/J\approx2-3$, and the
transition itself most likely describable within the Ginzburg-Landau paradigm.

The controversy arising from this difference in values of $Q_c/J$ was
recently addressed \cite{Kotov_2009} by Kotov {\it et al.}. By considering
effects of fluctuations around their trial (paramagnetic) columnar dimer state
(CDS), they came to the conclusion that a mean-field theory is incapable of
correctly describing the QPT in the $J$-$Q$ model. The nature and numerical
value of the phase transition point, found in their paper, was claimed to agree
with the work of Sandvik \cite{Sandvik_2007}. This circumstance was also used
to speculate that the small, compared to QMC, value of $Q_c/J$ resulting from
the analysis of Ref. \onlinecite{Isaev_2009}, is due to limitations of the HMF
approach. We would like to stress that the HMF method is not a {\it standard}
mean-field approach, as the one used in Ref. \onlinecite{Kotov_2009}, and it
becomes an exact method in the thermodynamic limit. Moreover, for a finite
system HMF can be implemented as a variational theory in terms of the energy,
and finite-size scaling needs to be performed to extrapolate to the
thermodynamic limit. 

In the present note we show that the claims of Ref. \onlinecite{Kotov_2009} are
misleading. We strongly oppose the statement ``Near the QCP [quantum critical
point], whose location [$K_c/J_K\approx2.16$] we find in fairly good agreement
with recent QMC studies...'', made in Ref. \onlinecite{Kotov_2009}, by
demonstrating that a direct comparison of the two results is inappropriate,
because the phase transition point, claimed by Kotov {\it et al.}, in reality
lies in the yet unexplored region $J<0$ of the $J$-$Q$ phase diagram. This
circumstance may raise doubts regarding the nature of the magnetic state found
by Kotov {\it et al.} We use a simple {\it variational} argument to show that
the large fluctuations found in their work could be attributed to an
instability of their reference dimer state against formation of a plaquette
condensate.

\begin{figure}[t]
 \begin{center}
  \includegraphics[width=0.9\columnwidth]{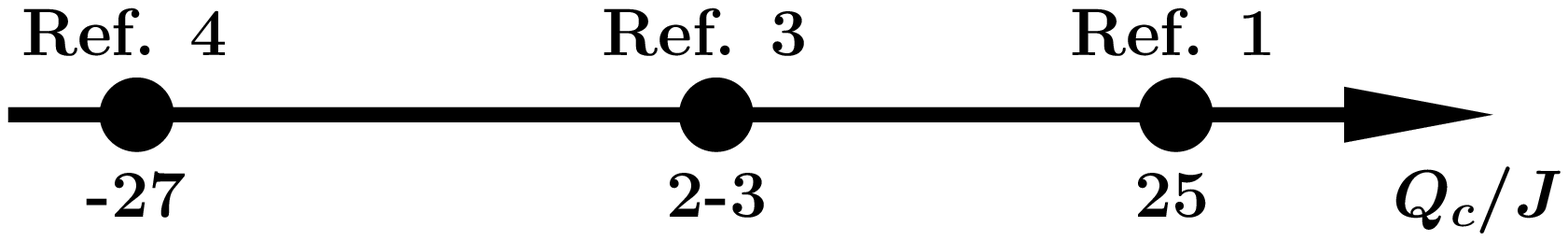}
 \end{center}
 \caption{$Q_c/J$ ($Q_c>0$), obtained in Refs. \onlinecite{Sandvik_2007},
          \onlinecite{Isaev_2009} and \onlinecite{Kotov_2009}.}
 \label{fig_status_quo}
\end{figure}

Up to an irrelevant constant, the Hamiltonian \eqref{jq_hamiltonian} can be
rewritten \cite{Kotov_2009} in the form:
\begin{equation}
 H_K=J_K{\textstyle\sum_{\langle ij\rangle}}\bS_i\bS_j-K{\textstyle
 \sum_{\langle ijkl\rangle}}(\bS_i
 \bS_j)(\bS_k\bS_l) 
 \label{JK}
\end{equation}
with $J_K,\,K>0$. The new coupling constant $K/J_K$ is related to the old one
$Q/J$ by the formula:
\begin{equation}
 Q/J=\bigl(K/J_K\bigr)/\bigl[1-K/2J_K\bigr]
 \label{new_representation}
\end{equation}
Clearly, for positive $Q$ and $J$ the parameter range that can be explored with
the Hamiltonian \eqref{JK} is $0\leqslant K/J_K\leqslant2$. The values of
$K/J_K$, larger than 2 correspond to the region $Q/J<0$, which implies either
(i) $Q<0$, $J>0$, or (ii) $Q>0$, $J<0$. We will assume, as Ref.
\onlinecite{Kotov_2009}, that $K$ and $Q$ have the same sign  and disregard the
case (i). The case (ii) defines the ferromagnetic (FM) part of the phase
diagram of the Hamiltonian \eqref{jq_hamiltonian}. In the new representation
\eqref{new_representation} the results of Refs. \onlinecite{Sandvik_2007} and
\onlinecite{Isaev_2009} are $K_c/J_K\approx1.85$ and $K_c/J_K\approx1-1.2$,
respectively. The critical value, obtained in Ref. \onlinecite{Kotov_2009}, is
$K_c/J_K\approx2.16$ which, after going back to the original units, corresponds
to $Q_c/J\approx-27$. This value should be compared with the result of QMC
simulations \cite{Sandvik_2007} $Q_c/J\approx+25$. Results of Refs.
\onlinecite{Sandvik_2007}, \onlinecite{Isaev_2009} and \onlinecite{Kotov_2009}
are summarized in Fig. \ref{fig_status_quo}. Thus, the claim \cite{Kotov_2009}
that the transition point $K_c/J_K\approx2.16$ is in ``good agreement'' with
the QMC result, is unjustified. A fair comparison involves more than
{\it numerology}, as we show below. 

\begin{figure}[t]
 \begin{center}
  \includegraphics[width=\columnwidth]{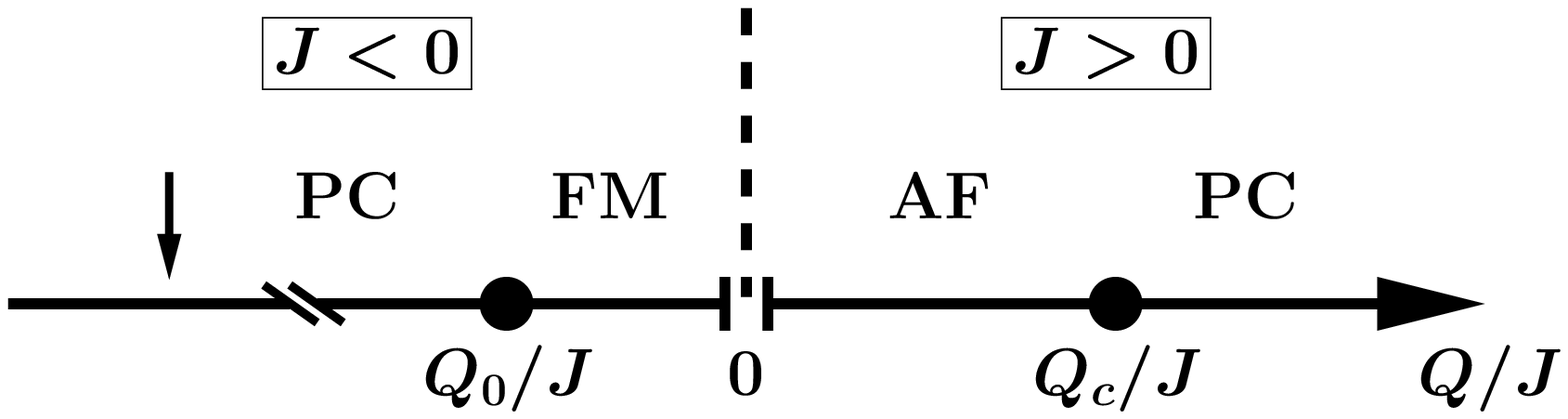}
 \end{center}
 \caption{Phase diagram of the $J$-$Q$ model for $Q>0$. The part $J>0$ was
          computed in Ref. \onlinecite{Isaev_2009}. The point
          $Q_0/J\approx-1.3$ indicates a $1^{\rm st}$ order QPT between the FM
          state and singlet plaquette crystal (PC) phase\cite{Isaev_2009}. The
          arrow shows $Q_c/J\approx-27$, obtained in Ref.
          \onlinecite{Kotov_2009}. The PC regions at $J<0$ and $J>0$ are
          adiabatically connected by changing $J$ ($Q/J$) through zero
          (infinity), i.e. there is no direct FM-to-AF transition.}
 \label{fig_phase_diagram}
\end{figure}

\begin{table}[b]
 \caption{Numerical values for the $1^{\rm st}$ order transition point $Q_0/J$,
          obtained using ED and HMF for different cluster sizes.}
 \begin{tabular}{c|c|c}
  & $2\times2$ & $4\times4$ \\
  \hline\hline
  ED & -1.0 & -1.19 \\
  \hline
  HMF & -1.32 & -1.26 \\
  \hline\hline
 \end{tabular}
 \label{tab_Q0}
\end{table}

Since the ratio $Q_c/J$ obtained by Kotov {\it et al.} is negative, one might
think that in reality Ref. \onlinecite{Kotov_2009} studies the FM-to-singlet
phase QPT. However, this is not necessarily the case. In fact, negative
values of $Q/J$ ($Q>0$) do not imply a FM phase. In order to address this
issue, we used the HMF approach \cite{Isaev_2009} and exact diagonalization
(ED), both in clusters of $2\times2$ and $4\times4$ spins, to determine the
$J$-$Q$ phase diagram in the AF ($J>0$) and FM ($J<0$) regimes. The results are
shown in Fig. \ref{fig_phase_diagram}. The QPT separating FM and singlet phases
is $1^{\rm st}$ order: it manifests itself as a level crossing both in ED and
HMF. In Table \ref{tab_Q0} we present numerical values for $Q_0/J$ obtained
from these methods. We see that they are in excellent agreement with each
other, which is not surprising, given the fact that the FM state is
semiclassical and the singlet phase is gapped. Thus, the paramagnetic phase
displays {\it two}, FM and AF instabilities. The region $K/J_K>2$ and the point
$K_c/J_K\approx2.16$ reside in the singlet phase of Fig.
\ref{fig_phase_diagram}, at negative values of $Q/J$. On the other hand, the
CDS \cite{Kotov_2009} can accomodate a N\'eel phase only if the magnetic unit
cell includes {\it two} dimers. Ref. \onlinecite{Kotov_2009} seems to consider
only homogeneous phases with one dimer per unit cell. This might raise doubts
regarding the AF nature of the CDS instability discussed by Kotov {\it et al.}

\begin{figure}[t]
 \begin{center}
  \includegraphics[width=\columnwidth]{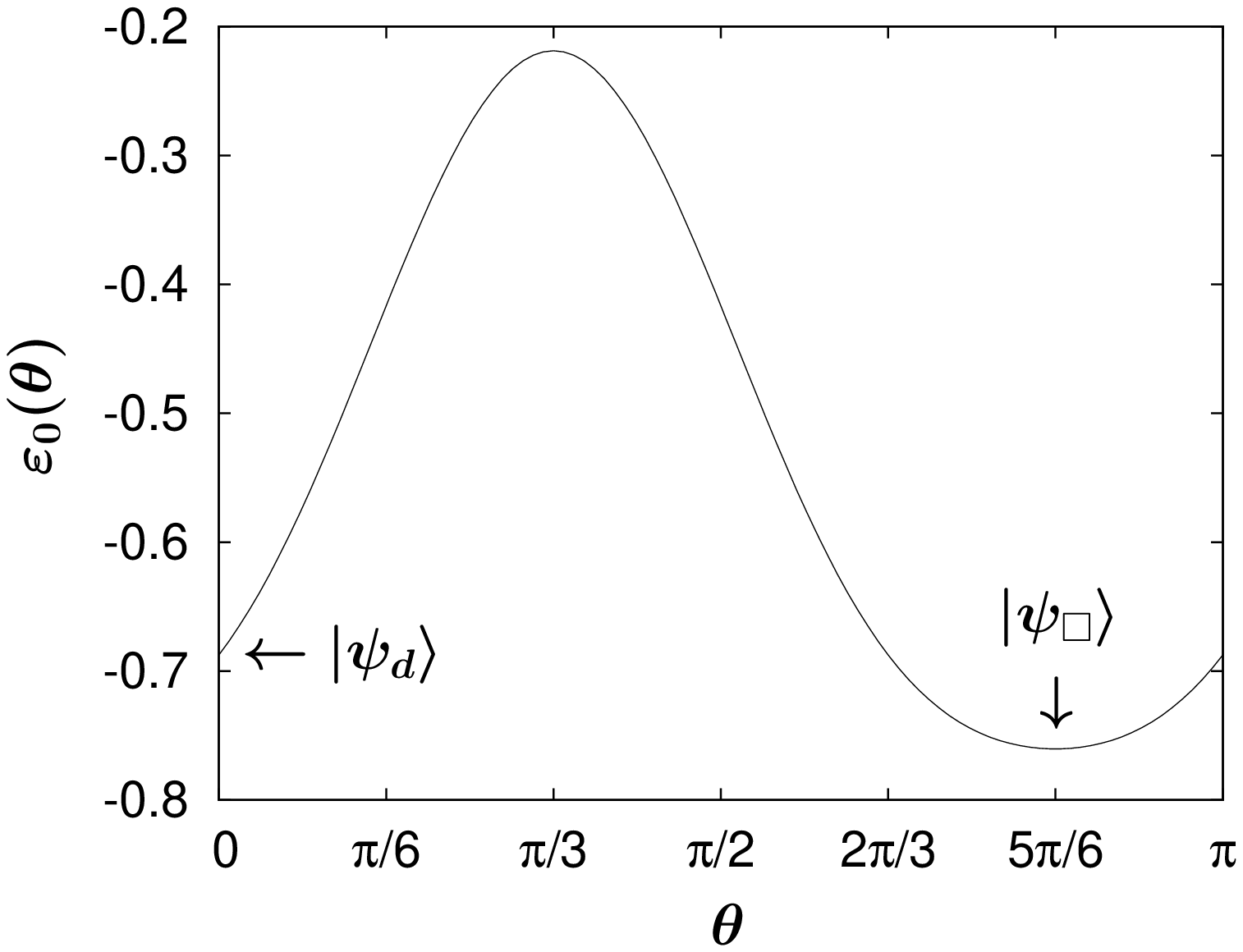}
 \end{center}
 \caption{Energy \eqref{trial_energy} as a function of $\theta$. The plaquette
          state corresponds to a local minimum $\varepsilon_0(5\pi/6)=-73/96$.
          The columnar dimer configuration has a higher energy
          $\varepsilon_0(0)=-66/96$ and does not correspond to an extremal
          point of $\varepsilon_0(\theta)$.}
 \label{fig_trial_energy}
\end{figure}

Finally, we shall comment on the huge fluctuation corrections to the value of
$K_c$, found in Ref. \onlinecite{Kotov_2009}. These corrections were argued to
be responsible for shifting the QPT towards negative values of $Q/J$ and seem
to be intimately related to the use by Kotov {\it et al.} of the CDS as a
physical vacuum for their analysis. This dimer state is unstable, when compared
to the plaquette structure \cite{Isaev_2009}. Let us consider a trial
paramagnetic state of the form:
\begin{equation}
 \vert\Psi_0\rangle=\prod_\Box\bigl[\cos\theta\vert\psi_d\rangle+
 \sin\theta\bigl(\vert\psi_\Box\rangle-\langle\psi_d\vert\psi_\Box\rangle\vert
 \psi_d\rangle\bigr)\bigr],
 \label{trial_state}
\end{equation}
where the product runs over $N/4$ $2\times2$ plaquettes. The wavefunction of
Eq. \eqref{trial_state} interpolates between the dimer condensate
\cite{Kotov_2009} $\prod\vert\psi_d\rangle$ ($\vert\psi_d\rangle$ is the direct
product of spin singlets on two parallel links of a plaquette), for $\theta=0$,
and the plaquette state \cite{Isaev_2009} $\prod\vert\psi_\Box\rangle$, which
corresponds to $\theta=5\pi/6$. The expectation value of Hamiltonian
\eqref{jq_hamiltonian}, $\varepsilon_0(\theta)\equiv\langle\Psi_0(\theta)\vert
H\vert\Psi_0(\theta)\rangle/NQ$, is given by:
\begin{align}
 \varepsilon_0&(\theta)=-\bigl(J/8Q\bigr)\bigl(\sin\theta-\sqrt{3}\cos\theta
 \bigr)^2-\bigl(1/96\bigr)\times \label{trial_energy} \\
 \times&\bigl(51+13\cos2\theta+2\cos4\theta-13\sqrt{3}\sin2\theta+2\sqrt{3}
 \sin4\theta\bigr). \nonumber
\end{align}
For $Q\gg J$, this function is shown in Fig. \ref{fig_trial_energy}. The
plaquette state corresponds to a local energy minimum. On the contrary, the CDS
does not describe any extremal point. Moreover, it has a higher energy compared
to the plaquette configuration. The energy difference between the CDS and a
correlated plaquette structure will be even larger if $4\times4$ spin clusters
\cite{Isaev_2009} are used as a basis for the HMF analysis.

JD acknowledges support from the Spanish DGI under Grant No. FIS2009-07277.

\end{document}